\documentclass[sigconf]{acmart}

\usepackage{booktabs} 
\usepackage{tabularx} 
\usepackage[english]{babel}
\usepackage{mathrsfs}

\usepackage{multirow}
\usepackage{multicol}
\usepackage{array}
\usepackage{arydshln}

\usepackage{amsmath}
\usepackage{subfig}

\usepackage{amssymb}
\usepackage{booktabs}
\usepackage[inline]{enumitem}

\newcounter{question}
\newcommand{\nextquestion}{\refstepcounter{question}\arabic{question}}
\usepackage[font=small,labelfont=bf]{caption}
\usepackage{algpseudocode}
\usepackage{algorithm}
\algnewcommand\algorithmicforeach{\textbf{for each}}
\algdef{S}[FOR]{ForEach}[1]{\algorithmicforeach\ #1\ \algorithmicdo\ }

\newcolumntype{L}[1]{>{\raggedright\arraybackslash}p{#1}}
\newcolumntype{C}[1]{>{\centering\arraybackslash}p{#1}}
\newcolumntype{R}[1]{>{\raggedleft\arraybackslash}p{#1}}

\newcommand{\mypar}[1]{\medskip\noindent\textbf{#1}~}

\author{Harshith Padigela, Hamed Zamani, and W. Bruce Croft}
\affiliation{%
  \institution{College of Information and Computer Sciences}
  \institution{University of Massachusetts Amherst}
 \city{Amherst} 
 \state{MA 01003}
}
\email{{hpadigela, zamani, croft}@cs.umass.edu}

\usepackage{xspace} 

\begin{document}

\title{Passage Re-ranking: A Comparative Analysis}
\title{Why BERT works? Case study on Passage Re-ranking}
\title{BERT for Passage Re-Ranking: Why Does It Work?}
\title{Investigating the Successes and Failures of BERT for Passage Re-Ranking}

\begin{abstract}
The bidirectional encoder representations from transformers (BERT) model has recently advanced the state-of-the-art in passage re-ranking. In this paper, we analyze the results produced by a fine-tuned BERT model to better understand the reasons behind such substantial improvements. To this aim, we focus on the MS MARCO passage re-ranking dataset and provide potential reasons for the successes and failures of BERT for retrieval. In more detail, we empirically study a set of hypotheses and provide additional analysis to explain the successful performance of BERT.

\end{abstract}
\maketitle

\section{Introduction}
Recent developments in deep learning and the availability of large-scale datasets have led to significant improvements in various computer vision and natural language processing tasks. In information retrieval~(IR), the lack of publicly available large-scale datasets for many tasks, such as ad-hoc retrieval, has restricted observing substantial improvements over traditional methods \cite{lin_rant,zamani2018sigir}. A number of approaches, such as weak supervision~\cite{dehghani2017neural,zamani2018theory}, have been recently proposed to enable deep neural models to learn from limited training data. More recently, Microsoft has released MS MARCO v2~\cite{ms_marco}, a large dataset for the passage re-ranking task, to foster the neural information retrieval research.

In this paper, we first show that a simple neural model that uses bidirectional encoder representations from Transformers (BERT)~\cite{bert} for question and passage representations performs surprisingly well compared to state-of-the-art retrieval models, including traditional term-matching models, conventional feature-based learning to rank models, and recent neural ranking models. This has been also discovered by other researchers, such as \cite{bert_marco} in parallel with this study. Looking at the leaderboard of the MS MARCO passage re-ranking task shows the effectiveness of the BERT representations for retrieval.\footnote{The leaderboard is available at \url{http://www.msmarco.org/leaders.aspx}.}

We believe that understanding the performance of effective neural IR models, e.g., BERT, is important. It could potentially provide guidelines for the IR researchers for further development of neural IR models. Given this motivation, this paper mainly analyzes the results obtained by BERT for passage re-ranking and studies the reasons behind its success. To do so, we compare the results obtained by both BM25 and BERT, and highlight their differences. We choose BM25 as our basis for comparison, due to its effectiveness and more importantly its simplicity and explainable behavior, which makes the analysis easier.

In more detail, this paper studies the following hypotheses: 
\begin{itemize}[leftmargin=*]
    \item H1: BM25 is more biased towards higher query term frequency compared to BERT.
    \item H2: Bias towards higher query term frequency hurts the BM25 performance.
    \item H3: BERT retrieves documents with more novel words.
    \item H4: BERT's improvement over BM25 is higher for longer queries.
\end{itemize}
In addition we also identify the query types for which BERT does and does not perform well. Our experiments provide interesting insights into the performance of this model.

\section{BERT}
Representations learned using language modelling \cite{mikolov2013distributed,glove} have shown to be useful in many downstream natural language tasks \cite{collobert2011natural}. There exist two primary approaches for using these pre-trained representations: (1) feature-based models and (2) fine-tuning \cite{bert}. In the feature-based approach, task-specific architectures are designed on top of the pre-trained feature representations. While in the fine-tuning approach, minimal task specific parameters are added, which will be fine-tuned in addition to the pre-trained representations for the downstream task. BERT \cite{bert}, which falls into the latter category, is a multi-layer bidirectional transformer encoder utilizing the transformer units described in \cite{transformer}. The BERT model uses bidirectional self-attention to capture interaction between the input tokens and is pre-trained on the masked language modelling task \cite{bert}.

Pre-trained BERT models, fine-tuned using a single additional layer have been shown to achieve state-of-the-art results in a wide range of natural language tasks, including machine reading comprehension (MRC) and natural language inference (NLI) \cite{bert}. In this paper, we also use the same setting, by adding a single layer on top of the BERT's large pre-trained model, and fine-tuning it using a pointwise setting with a maximum likelihood objective. This is also similar to the setting used in \cite{bert_marco}.


\section{Empirical Analysis}
\subsection{Data} \label{ssec:data}
We consider the MS MARCO dataset for passage re-ranking \cite{ms_marco} in our analysis. The MS MARCO dataset is generated using queries sampled from the Bing's query logs and corresponding relevant passages marked by human assessors. It is notable that the relevance judgments provided by the MS MARCO dataset are different from the traditional TREC-style relevance judgments. They utilized the information that the human assessors provided for the machine reading comprehension data. This means that a marked passage is a true positive/relevant, however an unmarked passage may not be a true negative. For every query, a set of top 1000 candidate passages are extracted using BM25 for re-ranking. Since the original relevant documents were picked from a set of 10 candidate documents chosen by the Bing's ranking stack, the relevant passages might not be present in the 1000 passages chosen by BM25.

The training set consists of approximately 400 million tuples of query, relevant passage, and non-relevant passage. The  development  set  contains   6,980  queries with their corresponding set of 1,000 candidate passages. On average, each query has one relevant passage. Around 1242 queries have no marked relevant passages. We primarily focus our analysis only on the 5738 queries in the development set that have at least one relevant passage.

\subsection{Experimental Setup}
We use BERT and BM25 for our analysis and comparison. We used the BERT large model trained on MS MARCO. The training setup is similar to the one described in \cite{bert_marco}. 
For BM25 relevance matching we indexed all the passages using Elasticsearch \cite{elasticsearch} with the default analyzer and default parameters of $b=0.75$ and $k1=1.2$. Since most queries have only 1 relevant document, we use mean reciprocal rank of the top 10 retrieved passages (MRR@10) as our main retrieval metric, which is also suggested by MS MARCO \cite{ms_marco}.\footnote{Due to the incomplete judgments, recall-oriented metrics such as mean average precision (MAP), are not suitable for this dataset.}

\subsection{Results and Discussion}
The performance of various models on the entire development and evaluation (test) sets of MS MARCO are shown in Table \ref{table:mrr}. We can see that the BERT model which was originally trained on the masked language modelling (MLM) task~\cite{bert} and further fine-tuned using a pointwise training on the MS MARCO data, outperforms the existing traditional retrieval models and recent neural ranking models by a large margin. In order to understand these improvements, we look into the BERT's and the BM25's performances on the development set.\footnote{Note that the evaluation set is not publicly accessible.} 

\begin{table}[t]
\caption{MRR@10 percentage from the MS MARCO \href{http://www.msmarco.org/leaders.aspx}{leaderboard}.}
\vspace{-0.2cm}
\begin{tabular}{lll}\toprule
\textbf{Model} & \textbf{Eval} & \textbf{Dev} \\\midrule
BM25  & 16.49  & 16.70   \\
BM25 (ours) & - & 17.67 \\
Feature-based LeToR: with RankSVM & 19.05 & 19.47 \\
Neural Kernel Match IR (KNRM)  & 19.82 & 21.84 \\
Neural Kernel Match IR (Conv-KNRM) & 27.12 & 29.02 \\
IRNet (Deep CNN/IR Hybrid Network) & 28.06 & 27.80 \\
\textbf{BERT + Small Training} & \textbf{35.87} & \textbf{36.53} \\ \bottomrule
\end{tabular}
\label{table:mrr}
\vspace{-4mm}
\end{table}



In the following, we study a set of hypotheses and provide empirical evidence to either validate or invalidate them.


\mypar{Hypothesis I: BM25 is more biased towards higher query term frequency compared to BERT.} 
We hypothesize that in many queries the top results from BM25 were just the passages that contain multiple repetitions of query words without actually conveying any useful information, which is not the case for BERT. 
To validate this hypothesis, we calculate the \textit{fraction of query tokens (FQT)} as follows: For each query, we take the top $k$ results, remove stopwords and punctuations, and calculate the fraction of query tokens in the remaining tokens. If $d_1, d_2, \cdots, d_{k}$ are the set of results for a query $q$ without stopwords and punctuation, then,
\begin{equation}
  FQT(q) = \frac{1}{k}\sum_{i=1}^{k} \frac{N(d_i, q)}{|d_i|}  
\end{equation}
where $N(d_i, q)$ denotes the number of occurrences of query tokens $q$ in the document $d_i$. We limit $k$ to a maximum of 10.

We find that the FQT average across queries is 0.2 for BM25 and 0.147 for BERT. In 95.96\% of the queries BM25 has a higher FQT value than BERT. These results validate our first hypothesis, saying that BM25 has a higher bias towards query term frequency in document matches, compared to BERT. An example can be seen in Table \ref{table:obs} Query \ref{q:fqt}



\mypar{Hypothesis II: Bias towards higher query term frequency hurts the BM25 performance.} 
We hypothesize that the bias towards query term frequency affects the BM25 performance significantly, compared to BERT.


To investigate this, we see how MRR changes across different ranges of FQT. The FQT range of [0,1] is split into 5 buckets and the average MRR value and the number of queries in each bucket is shown in Table \ref{table:qtf_mrr_compare}. 
As FQT value increases, we can see that the MRR value decreases in both BERT and BM25. Because of BM25's bias towards high FQT (validated by Hypothesis I and also evident by the number of queries), we can see the decrease in MRR (as we go from the lowest to highest FQT buckets) is more prominent for BM25 (34.5\%)  than BERT (16.7\%). The signed t-test for measuring the difference between two pairs of data, applied on difference between FQT values of BM25 and BERT yields a p-value of 0.0, indicating statistically significant difference between the FQT values.


\begin{table}[t]
\caption{Average MRR and \# of queries (in parenthesis) for ranges of FQT.}
\vspace{-0.2cm}
\begin{tabular}{l|l|l|l|l|l}\toprule
  \textbf{FQT} & $[0, 0.1)$ & $[0.1, 0.15)$ & $[0.15, 0.2)$ & $[0.2, 0.25)$ & $[0.25, 1]$  \\\midrule
\textbf{BM25} & 0.29 & 0.23 & 0.22 & 0.20 & 0.19\\
              & (349) & (1163) & (1565) & (1316) & (1345)\\
\textbf{BERT} & 0.48 & 0.47 & 0.42 & 0.38 & 0.40\\ 
              & (1240) & (2061) & (1441) & (652) & (344) \\
\bottomrule
\end{tabular}
\label{table:qtf_mrr_compare}
\end{table}


\begin{table}[t]
\centering
\caption{Average MRR with respect to query length (L).}
\vspace{-0.2cm}
\resizebox{1.04\columnwidth}{!}{%
    \begin{tabular}{l|l|l|l|l|l|l|l|l|l}\toprule
    \textbf{L} & 2 & 3    & 4    & 5   & 6    & 7    & 8    & 9    & 10   \\
    \hline
    \textbf{BM25}  & 0.27 & 0.23 & 0.22 & 0.22& 0.23 & 0.19 & 0.21 & 0.17 & 0.18 \\
    \textbf{BERT} & 0.56 & 0.46 & 0.48 & 0.45 & 0.46 & 0.42 & 0.40 & 0.38 & 0.34\\
    \bottomrule
    \end{tabular}
}
\label{table:mrr_qlen}
\vspace{-3mm}
\end{table}

\mypar{Hypothesis III: BERT retrieves documents with more novel words.} 
Since the recent neural models trained on the language modeling task have been shown to capture semantic similarities, we hypothesize that BERT can retrieve results with more novel words, compared to BM25. To validate this, we calculate the \textit{fraction of novel terms (FNT)} as follows. Let $d_1, d_2, \cdots, d_{k}$ be the results for a query $q$, which are stripped of stopwords and punctuation. Then 
\begin{equation}
  FNT(q) = \frac{1}{k}\sum_{i=1}^{k} \frac{N'(d_i, q)}{U(d_i)}  
\end{equation}
where $U(d_i)$ gives the number of unique terms in document $d_i$ and $N'(d_i, q)$ gives the number of unique terms in document $d_i$ which are not present in the query $q$. We limit $k$ to a maximum of 10. We find that the FNT average across queries is 0.88 for BM25 and 0.9 for BERT. In $85.85\%$ of queries BERT has a higher FNT value than BM25. The signed t-test on the difference between FNT values of BERT and BM25 yields a p-value of 0.0, indicating statistically significant difference between the FNT values. This validates our hypothesis that BERT retrieves documents with more novel words than BM25. 


\begin{table}[t]
\centering
\caption{Average MUR with respect to the different cut-off values (i).}
\vspace{-0.2cm}
\resizebox{1.04\columnwidth}{!}{%
    \begin{tabular}{l|l|l|l|l|l|l|l|l|l|l}\toprule
    \textbf{i} & 1    & 2    & 3    & 4   & 5    & 6    & 7    & 8    & 9    & 10   \\
    \hline
    \begin{tabular}[c]{@{}l@{}}\textbf{Avg}.\\ \textbf{MUR}\end{tabular}  & 0.17 & 0.45 & 0.77 & 1.1 & 1.44 & 1.78 & 2.13 & 2.46 & 2.8 & 3.12\\
    \bottomrule
    \end{tabular}
}
\label{table:mur}
\vspace{-0.2cm}
\end{table}

\mypar{Hypothesis IV: BERT$'$s improvement over BM25 is higher for longer queries.}
Since the BERT model is designed to learn context-aware word representations, we hypothesize that its improvements for longer queries, which generally provide richer context, are more significant. To validate this hypothesis, we calculate the average MRR per query length for both BERT and BM25, shown in  Table~\ref{table:mrr_qlen}. We can see that BERT performs significantly better than BM25 across all query lengths. But as the query length increases from 2 to 10, the performance of both BM25 and BERT generally decreases, 
and this decrease is more prominent for BERT (39\%) compared to BM25 (33\%), indicating its higher sensitivity to query length than BM25. The MRR difference between BERT and BM25 also decreases from 0.29 to 0.16 as query length increases from 2 to 10, which indicates that our fourth hypothesis is incorrect and that BERT's improvement is lower for longer queries. Interestingly, BERT performs surprisingly well for very short queries compared to longer ones. The reason might be that BERT is not  successful at capturing the query context properly for long queries. This can be also observed from the examples, such as Queries \ref{q:long}, \ref{q:specific} in Table~\ref{table:obs}.



\begin{table*}[t]
  \caption{Sample queries for comparison. (W) - incorrect and (C) - correct result.}
  \vspace{-3mm}
  \begin{tabular}{c|l|l|c}\toprule
   \textbf{ID} & \textbf{Query} & \textbf{BM25/BERT/Relevant passage} & \textbf{Comparison} \\
    \hline
    \nextquestion\label{q:fqt} & what is the nationality & \begin{tabular}[c]{@{}l@{}}\textit{BM25}: Users found this page by searching for 1 is african american a nationality \\ 2 african american nationality 3 is black a nationality  nationality african ....\\\textit{BERT}: Nationality is the legal relationship between a person and a nation state ....\end{tabular} & \begin{tabular}[c]{@{}l@{}}BM25(W) vs\\ BERT(C)\end{tabular} \\
    \hline 
    \nextquestion\label{q:bow} &  confident man definition & \begin{tabular}[c]{@{}l@{}}\textit{BM25}: definition of suave is someone smooth confident .... usually describing a man \\ \textit{BERT}: definition of a confidence man is someone who gets a victim to trust them \\ before taking their money or property, a con man .... \end{tabular} & \begin{tabular}[c]{@{}l@{}}BM25(W) vs\\ BERT(C)\end{tabular}\\
    \hline 
    \nextquestion\label{q:long} & \begin{tabular}[c]{@{}l@{}}where can a plasma mem- \\ brane be found in a cell\\ \end{tabular} & \begin{tabular}[c]{@{}l@{}} \textit{BERT}: The Plasma membrane is found in both the animal cell and plant cell\\ \textit{Rel}: The plasma membrane is the border between the interior and exterior of a cell ....\end{tabular} & \begin{tabular}[c]{@{}l@{}} BERT(W) vs\\ Relevant\end{tabular} \\
    \hline
    \nextquestion\label{q:specific} & \begin{tabular}[c]{@{}l@{}}telephone number for amazon \\ fire stick customer service\end{tabular} &  \begin{tabular}[c]{@{}l@{}}\textit{BERT}: Customer Service  1 866 216 1072. .... 1 Thank you for calling Amazon.com \\ customer service for quality assurance and training .... \\ \textit{Rel}: .... for more information contact amazon fire stick support number 1 8447451521\end{tabular} & \begin{tabular}[c]{@{}l@{}} BERT(W) vs\\ Relevant\end{tabular}\\
    \hline
    \nextquestion\label{q:context} & another name for reaper & \begin{tabular}[c]{@{}l@{}}\textit{BERT}: Reaper  orginally known as Gabriel Reyes  is a mercenary.... is antagonist .... in\\ videogame Overwatch. He is voiced by Keith Ferguson who also played Lord Hater \\ \textit{Rel}: .... similar words for the term reaper. harvester reaper ....\end{tabular} &  \begin{tabular}[c]{@{}l@{}} BERT(W) vs\\ Relevant\end{tabular}\\
    \bottomrule
  \end{tabular}
  \label{table:obs}
\end{table*}

\subsection{Result Analysis}
We conduct various analyses to understand the similarities and differences between BM25 and BERT. We discuss them below.

\begin{figure}[t]
\includegraphics[width=0.6\linewidth]{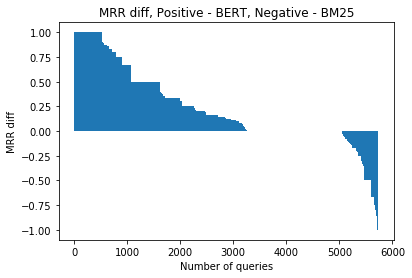}
\vspace{-0.2cm}
\caption{$\text{MRR}_{\text{BERT}} - \text{MRR}_{\text{BM25}}$ on MSMarco Dev set.}
\vspace{-0.2cm}
\label{fig:mrr_diff}
\end{figure}

\mypar{Per Query Analysis.} We analyze the per query performance of BERT compared to BM25. Figure \ref{fig:mrr_diff} plots $\Delta$MRR per query (i.e., $\text{MRR}_{\text{BERT}} - \text{MRR}_{\text{BM25}}$), which are sorted in descending order. As depicted in the figure, in 3257 questions (57\% out of 5738) BERT has a better performance compared to BM25 and in 690 questions (12\%) BM25 performs better than BERT. For $525$ (9\%) queries $\Delta$MRR is equal to $1$, meaning that a relevant answer is retrieved by BERT as the first ranked passage, however, no relevant answer is retrieved by BM25 in the top 10 result list. For 1791 queries, both BERT and BM25 perform similarly. 

This experiments show that BERT not only performs better than BM25 (on average), but it also performs more accurately for substantially more queries. 

\mypar{Similarity between the BERT's and the BM25's result lists.}
To measure the similarity between the results of BERT and BM25, we calculate the following metric, MUR - matches upto result. MUR$(i,q)$ for a query $q$ measures the number of matches in the top $i$ results of BERT and BM25. 
We can see the average MUR for each $i \in [1,10]$ in Table \ref{table:mur} which indicates the low extent of similarity between BERT and BM25. The number of matches increases linearly with $i$ with slope of about $0.33$ and intercept around -$0.21$, indicating a consistent linear relationship between BERT and BM25.


\mypar{Comparison by answer type.} In order to understand the performance of these models across different types of questions, we classify questions based on the lexical answer type. We use the rule-based answer type classifier\footnote{\url{https://github.com/superscriptjs/qtypes}} inspired by \cite{li2002learning} to extract answer types. We classify questions based on 6 answer types, namely abbreviation, location, description, human, numerical and entity. The average MRR across these 6 types for 4105 queries (having a valid answer type) is shown in Table \ref{table:ans_type}. We can see that while BERT has highest MRR on \textit{abbreviation} type questions, BM25 has its lowest MRR on them. Note that BERT seems to have its lowest performance on \textit{numerical} and \textit{entity} type questions.

\begin{table}[t]
\caption{Average MRR values for answer types. Sorted by $\Delta$MRR.}
\vspace{-0.2cm}
\begin{tabular}{l|l|l|l|l|l|l}\toprule
\textbf{Type} & ABBR & LOC & DESC & HUM & NUM & ENTY  \\
\hline
\textbf{\# queries} & 9 & 493 & 1887 & 455 & 933 & 328 \\ 
\textbf{BM25} & 0.17 & 0.25 & 0.19 & 0.23 & 0.19 & 0.21  \\
\textbf{BERT} & 0.59 & 0.50 & 0.43 & 0.46 & 0.40 & 0.41 \\ \bottomrule
\end{tabular}
\label{table:ans_type}
\vspace{-3mm}
\end{table}

\begin{figure}[t]
    \centering
    \includegraphics[width=0.9\linewidth]{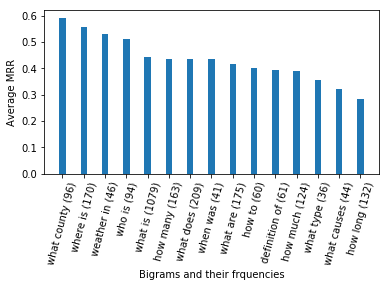}
    \vspace{-0.2cm}
    \caption{Average MRR of Frequent Bigrams.}
    \label{fig:bigram_mrr}
\end{figure}

\mypar{Comparison using query starting ngrams:} Here we look at the most frequent bigrams with which the queries start. The idea is that looking at the starting ngrams can help us understand the type of queries. We extract the most frequent 15 bigrams and compute the average MRR using BERT for each of them. The result is shown in Figure~\ref{fig:bigram_mrr}. We can see that the bigrams corresponding to \textit{numeric} type questions, such as ``how much'' and ``how long'', as well as \textit{location} type questions like ``what county / where is'' and \textit{entity} type questions, such as ``what type'' have a low MRR. This is consistent with our observations in the previous experiment (see Table \ref{table:ans_type}).

\begin{figure}[t]
    \centering
    \includegraphics[width=0.8\linewidth]{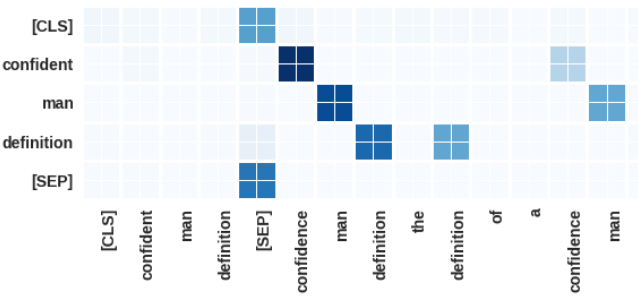}
    \vspace{-0.2cm}
    \caption{Attention map of head 14 from BERT layer 16.}
    \label{fig:pos_attn}
\end{figure}

\begin{figure}[t]
    \centering
    \includegraphics[width=0.8\linewidth]{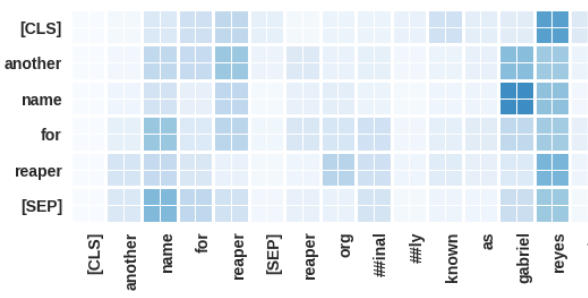}
    \vspace{-0.2cm}
    \caption{Attention map of head 4 from BERT layer 23.}
    \label{fig:neg_attn}
    \vspace{-0.2cm}
\end{figure}

\mypar{Semantic similarity:}
Being trained on a language modeling task, we expect BERT to capture various semantic relationships. While in some cases these help in arriving at the right answer sometimes they can also lead to incorrect answers. We will discuss two such examples below. In question \ref{q:bow} of Table~\ref{table:obs}, BERT captures similarity between the word ``confident'' in query and ``confidence'' in the passage, which helps it arrive at the right answer. This can be seen by visualizing the attention values between query and document words as shown in Figure \ref{fig:pos_attn}. Similarly in Example \ref{q:context} of Table~\ref{table:obs}, the question asks for another \textit{name} for word ``reaper'', which in this context means synonyms for the word ``reaper''. However, BERT relates \textit{name} to a \textit{character name} reaper (see attention map \ref{fig:neg_attn}). This leads to an incorrect answer. 

\section{Conclusions and Future Work}
BERT performs surprisingly well for a passage re-ranking task. In this paper, we provide empirical analysis to understand the performance of BERT and how its results are different from a typical retrieval model, e.g., BM25. We showed that BM25 is more biased towards high query term frequency and this bias hurts its performance. We demonstrated that, as expected, BERT retrieves passages with more novel words. Surprisingly, we found out that BERT is failing at capturing the query context for long queries. Our analysis also suggested that BERT is relatively successful in answering \textit{abbreviation} answer type questions and relatively poor at \textit{numerical} and \textit{entity} type questions. 


Although BERT substantially outperforms state-of-the-art models for passage retrieval, it is still far away from a perfect retrieval performance. We believe that future work investigating the relevance preferences captured by BERT across various query types and a better encoding of query context for longer queries could help in developing even better models.

\vspace{-1mm}

\section{Acknowledgements}
This work was supported in part by the Center for Intelligent Information Retrieval and in part by NSF IIS-1715095. Any opinions, findings and conclusions or recommendations expressed in this material are those of the authors and do not necessarily reflect those of the sponsor.

\end{document}